\documentclass[aps,prd,showpacs,preprint]{revtex4}
\usepackage[british]{babel}

\usepackage{amsmath}
\usepackage{amssymb}

\usepackage{bbm}

\DeclareFontFamily{OT1}{rsfs}{}
\DeclareFontShape{OT1}{rsfs}{m}{n}{ <-7> rsfs5 <7-10> rsfs7 <10->rsfs10}{} 
\DeclareMathAlphabet{\mycal}{OT1}{rsfs}{m}{n}

\newcommand{\dual}[1]{\overset{{}^{{}^{\boldsymbol{\neg}}}}{\smash[t]{#1}}}
\newcommand{\M}{M_{\rm pl}}
\newcommand{\gdual}[1]{\overset{\:{}^{{}^{\boldsymbol{\neg}}}}{\smash[t]{#1}}}

\begin{document}

\title{Dark spinor inflation -- theory primer and dynamics}

\author{Christian G. B\"ohmer}
\affiliation{Department of Mathematics, University College London, 
             Gower Street, London, WC1E 6BT, United Kingdom}
\email{c.boehmer@ucl.ac.uk}

\date{\today}

\begin{abstract}
Inflation driven by a single dark spinor field is discussed. We define the notion of a dark spinor field and derive the cosmological field equations for such a matter source. The conditions for inflation are determined and an exactly solvable model is presented. We find the power spectrum of the quantum fluctuation of this field and compare the results with scalar field inflation.
\end{abstract}


\maketitle

\section{Introduction}

Inflation is a successful theory to solve some of the problems of the otherwise successful big-bang model~\cite{Guth:1980zm,Linde:1981mu,Albrecht:1982wi}, see also~\cite{Hollands:2002yb}. Today's standard model of cosmology contains the theory of scalar field inflation where the quantum fluctuations yield the primordial density fluctuations that become the seeds of structure formation. In this paper, inflation driven by a single dark spinor field is investigated. We present the basic theory of dark spinor inflation and formulate the theory in closest possible analogy to scalar field inflation. 

Spinors and inflation have not been studied to such a great extent in the past, however, see for instance~\cite{ArmendarizPicon:2003qk,Saha:2006iu,Boehmer:2007ut}. Spin and inflation in the context of torsion theories on the other hand has received much more attention~\cite{Gasperini:1986mv,Fennelly:1988dx,Chatterjee:1993rc,Obukhov:1993fd,Kao:1993nb,GarciadeAndrade:1999qt,Boehmer:2005sw}. However, here we consider spinors in the framework of classical general relativity. Also vector field driven inflation has attracted some interest~\cite{Ford:1989me,Burd:1991ew,Lidsey:1991ex,Lewis:1991nj,Golovnev:2008cf}.

The dark spinor field is a spin one half matter field with mass dimension one~\cite{jcap,prd}, originally called Elko spinor. The Elko spinors are based on the eigenspinors of the charge conjugation operator, the field theory obeys the unusual property $(CPT)^2=-\mathbbm{1}$. The spinor field's dominant interactions take place through the Higgs doublet or with gravity~\cite{jcap,prd}, and therefore these spinors are genuinely dark. Henceforth we will refer to dark spinors. They belong to a wider class of so-called flagpole spinors~\cite{daRocha:2005ti}. Physical properties and their relationship to Dirac spinors have been studied in~\cite{daRocha:2007pz}. According to Wigner~\cite{Wigner:1939cj}, these spinors are non-standard, standard fields obey $(CPT)^2=\mathbbm{1}$. First results about the imprints, a non-standard spinor driven inflation would leave on the cosmic microwave background anisotropies have been obtained in~\cite{Boehmer:2007ut}, in the context of Bianchi type $I$ models. 

Consider the left-handed part $\phi_L$ of a Dirac spinor in Weyl representation. An eigenspinor of the charge conjugation operator is defined by 
\begin{equation}
  \lambda = 
  \left(\begin{array}{c} 
    \pm \sigma_2 \phi^{\ast}_{L} \\
    \phi_L 
  \end{array}\right) \,,
  \label{eq:n1}
\end{equation}
where $\phi^{\ast}_{L}$ denotes the complex conjugate of $\phi_L$ and $\sigma_2$ is the second Pauli matrix. Note that the helicities of $\phi_L$ and $\sigma^2 \phi^{\ast}_L$ are opposite~\cite{jcap}. Therefore there are two distinct helicity configurations denoted by $\lambda_{\{-,+\}}$ and $\lambda_{\{+,-\}}$.

Dark spinor fields can be introduced straightforwardly into an arbitrary manifold~\cite{Boehmer:2006qq,Boehmer:2007dh}. This results in the following matter action
\begin{equation}
      S = 
      \frac{1}{2} \int \Bigl(g^{\mu\nu}
      \nabla_{(\mu} \dual{\lambda} \nabla_{\nu)} \lambda
      -V(\dual{\lambda} \lambda) \Bigr) \sqrt{-g}\, d^4 x\,,
      \label{eq:d14}
\end{equation}
where $V$ is the potential and the round brackets denote symmetrization. Since the covariant derivative now acts on a spinor, there is an additional coupling of the matter to the spin connection which is absent in the scalar field case.

In view of the dark spinor action~(\ref{eq:d14}) these fields obey the Klein-Gordon equation only. Dirac spinors, on the other hand, satisfy an additional equation of motion, namely the Dirac equation which is linear in the partial derivatives. Therefore, dark spinors show a more involved coupling to geometry than Dirac spinors. The other key difference between Dirac and dark spinors is their transformation behavior under the action of $(CPT)^2$. Dirac spinors satisfy $(CPT)^2=\mathbbm{1}$ while dark spinors satisfy $(CPT)^2=-\mathbbm{1}$.

In cosmology we assume the dark spinors to depend on time only. Following~\cite{Boehmer:2007dh} the dark spinors are given by
\begin{equation}
      \lambda_{\{-,+\}} = \varphi(t)\, \xi /\sqrt{2}\,,\qquad
      \lambda_{\{+,-\}} = \varphi(t)\, \zeta /\sqrt{2}\,,
      \label{eq:dy24}
\end{equation}
where $\xi$ and $\zeta$ are constant spinors~\cite{Boehmer:2007dh} satisfying $\gdual{\xi} \xi = \gdual{\zeta} \zeta = + 2$. $\gdual{\xi}$ and $\gdual{\zeta}$ denote the respective dual spinors. Note that these constant spinors are the only ones compatible with homogeneity and isotropy in the non-standard spinor context~\cite{Boehmer:2007dh}. 

The paper in organized in the following manner: In Section~\ref{sec:cos} the cosmological field equations of dark spinor inflation are formulated and compared with scalar field inflation. In Section~\ref{sec:spin} the theory of dark spinor inflation is formulated. The slow-roll approximations are applied to the equations of motion and a new slow-roll parameter is introduced. An exact inflationary solution is described in Section~\ref{sec:exact}. An exponential inflation model is derived. In Section~\ref{sec:pert} the power spectrum of the spinor field perturbations is found and compared with the scalar field analog. In the final Section~\ref{sec:concl} we conclude.

\section{Cosmological field equations}
\label{sec:cos}

The standard model of cosmology is based upon the flat Friedmann-Lema\^{\i}tre-Robertson-Walker (FLRW) metric
\begin{equation}
  ds^2 = dt^2 - a(t)^2 (dx^2+dy^2+dz^2),
  \label{flrw}
\end{equation}
where $a$ is the scale factor. The dynamical behavior of the universe is determined by the cosmological field equations of general relativity 
\begin{equation}
  R_{\alpha\beta} - \frac{1}{2}R g_{\alpha\beta} + \Lambda g_{\alpha\beta} = \frac{1}{\M^2} T_{\alpha\beta},
  \label{eq1}
\end{equation}
where $\M$ is the Planck mass which we use as the coupling constant. In what follows the cosmological constant is set to zero unless stated otherwise. $T_{\alpha\beta}$ denotes the stress-energy tensor, which for a homogeneous and isotropic cosmology takes the form
\begin{equation}
  T_{\alpha\beta} = \mbox{diag}(\rho,a^2 P,a^2 P,a^2 P).
  \label{eq2}
\end{equation}
The cosmological equations of motion are
\begin{eqnarray}
  H^2 = \frac{1}{3\M^2}\rho,
  \label{eq5}\\
  \dot{\rho} + 3H(\rho + P) = 0.
  \label{eq7}
\end{eqnarray}
where the dot denotes differentiation with respect to cosmological time $t$. The Hubble parameter $H$ is defined by $H=\dot{a}/a$.

Here we consider a homogeneous single dark spinor field $\lambda$, which we will call spinflaton. Note that this is different from spinflation in the context of DBI brane world models~\cite{Easson:2007dh}. For scalar field inflation, we refer the reader to~\cite{Liddle:2000,Dodelson:2003,Bassett:2005xm}. Following~\cite{Boehmer:2007dh}, the energy density and the pressure of the dark spinor field are given by
\begin{eqnarray}
  \rho = \frac{1}{2}\dot{\varphi}^2 + V(\varphi) - \frac{3}{8}H^2\varphi^2,
  \label{eq8}\\
  P = \frac{1}{2}\dot{\varphi}^2 - V(\varphi) + \frac{1}{8}H^2\varphi^2.
  \label{eq9}
\end{eqnarray}
The spinor field's potential energy may yield an accelerated expansion of the universe. It should be noted that the energy density and the pressure now explicitely depend on the Hubble parameter. These additional terms are present because the covariant derivative when acting on the spinflaton field acts also on the spinorial part. One way of interpreting this `coupling' is to regard the effective mass of the particle to depend on the Hubble parameter~\cite{Boehmer:2007dh} and therefore on the evolution of the universe.

The time-time component of the Einstein field equation~(\ref{eq5}) and the energy momentum conservation equation~(\ref{eq7}) of the spinflaton field become
\begin{eqnarray}
  H^2 = \frac{1}{3\M^2}\Bigl[\frac{1}{2}\dot{\varphi}^2 + V(\varphi) - \frac{3}{8}H^2\varphi^2\Bigr],
  \label{eq10}\\
  \ddot{\varphi} + 3H\dot{\varphi}\Bigl[1 - \frac{1}{4}\frac{\varphi}{\dot{\varphi}}\bigl(H+\frac{\ddot{a}}{a}\frac{\varphi}{\dot{\varphi}}\bigr)\Bigr] + V_{\varphi}(\varphi) = 0,
  \label{eq11}
\end{eqnarray}
where $V_{\varphi} = dV/d\varphi$. The additional term with the Hubble parameter on the right-hand side of Eq.~(\ref{eq10}), and the additional terms in the square bracket of Eq.~(\ref{eq11}) characterize the deviations from the scalar field equations where these terms are absent. Consequently, the theory of spinflation may be quite different to scalar field inflation. 

\section{Spinflation}
\label{sec:spin}

The standard model of cosmology is plagued by the fact that the universe's expansion is decelerating which leads to the flatness and horizon problems, among others. These problems can be solved if the universe did undergo a phase of accelerated expansion in the early universe~\cite{Guth:1980zm,Linde:1981mu,Albrecht:1982wi}, this means $\ddot{a} > 0$. This condition is equivalent to the requirement $\rho + 3P < 0$ which violates the strong energy condition. Taking into account the energy density and the pressure of the spinflaton field, one finds
\begin{equation}
  \frac{\rho + 3P}{2} = \dot{\varphi}^2 - V(\varphi),
  \label{sp1}
\end{equation}
and therefore the condition for the spinflaton field to yield a phase of accelerated expansion is given by
\begin{equation}
  V(\phi) > \dot{\varphi}^2.
  \label{sp2}
\end{equation}
This latter condition equals the analog condition of the scalar field case. It means physically that the potential energy of the spinflaton dominates its kinetic energy. This can be achieved quite generally if a sufficiently flat potential is considered where the spinflaton slowly rolls down. 

Next, let us impose slow-roll conditions: By this we mean that the kinetic energy of the spinflaton is much smaller than its potential energy $\dot{\varphi}^2/2 \ll V(\varphi)$ and that the `acceleration' of the field is small $|\ddot{\varphi}| \ll 3H|\dot{\varphi}|$. Then, the equations of motion~(\ref{eq10}) and~(\ref{eq11}) are approximately given by
\begin{eqnarray}
  H^2 \simeq \frac{1}{3\M^2}\Bigl[V(\varphi) - \frac{3}{8}H^2\varphi^2\Bigr],
  \label{sp4}\\
  3H\dot{\varphi}\Bigl[1 - \frac{1}{4}\frac{\varphi}{\dot{\varphi}}\bigl(H-\frac{\ddot{a}}{a}\frac{\varphi}{\dot{\varphi}}\bigr)\Bigr] \simeq -V_{\varphi}(\varphi),
  \label{sp5}
\end{eqnarray}
where the $\simeq$ sign indicates equality within the slow-roll approximation. At this stage it is not straightforward to introduce slow-roll parameters that can be expressed entirely in terms of the potential.

It is tempting to replace $H$ and $\ddot{a}/a$ in Eqs.~(\ref{sp4}) and (\ref{sp5}) by using the field equations again, thereby adding terms proportional to the inverse Planck mass. The resulting equations can be expanded in terms of powers of $\varphi/\M$, which leads to 
\begin{eqnarray}
  H^2 \simeq \frac{1}{3\M^2}V(\varphi)\Bigl[1 - \frac{1}{8}\frac{\varphi^2}{\M^2}\Bigr] + O(\varphi^5/\M^5),
  \label{sp6}\\
  3H\dot{\varphi} \simeq -V_{\varphi}(\varphi)\Bigl[1 - \frac{1}{8}\frac{\varphi^2}{\M^2} - \frac{1}{4\M}\frac{V(\varphi)}{V'(\varphi)}\frac{\varphi}{\M}\Bigr] + O(\varphi^3/\M^3),
  \label{sp7}
\end{eqnarray}
In order to investigate which simplifications are in fact justified, we will analyze some exact solutions of the non-approximated equations~(\ref{eq10})--(\ref{eq11}) in Section~\ref{sec:exact}. 

To quantify the amount of inflation, we define the number of $e$-foldings 
\begin{equation}
  N(t) = \ln \frac{a(t_{\rm end})}{a(t)},
\end{equation}
which measures the amount of inflation that still has to occur until inflation ends. By definition $N(t_{\rm end}) = 0$. Using the slow-roll approximation, we can express the number of $e$-foldings in terms of the potential
\begin{equation}
  N(t) = \int_t^{t_{\rm end}} H(t') dt' \simeq \frac{1}{\M^2}
  \int_{\varphi_{\rm end}}^\varphi \left[\frac{V}{V'}+
  \frac{1}{4}\frac{\varphi}{\M^2}\frac{V^2}{V'^2} \right]d\varphi,
\end{equation}
where higher order terms have been neglected. Throughout this paper we assume the total number of $e$-foldings to be around sixty.

The standard slow-roll parameter $\mathcal{E}$ is defined to be the time derivative of the inverse Hubble parameter. In principle, using Eqs.~(\ref{sp4}) and~(\ref{sp5}), we can express $(1/H)\dot{}$ in terms of the potential $V$, its derivative and the field $\varphi$. The exact expression is quite involved. However, by expanding the equation in terms of powers of $\varphi/\M$ one arrives at
\begin{equation}
  \mathcal{E} = -\frac{\dot{H}}{H^2} \simeq \frac{\M^2}{2}\Bigl(\frac{V'}{V}\Bigr)^2 - \frac{\varphi}{4}\frac{V'}{V}.
  \label{sp9}
\end{equation}
Let us next perform the analog computation for the complementary parameter $\Delta$ which also quantifies the slow-roll approximation
\begin{equation}
  \Delta = \frac{1}{H}\frac{\ddot{\varphi}}{\dot{\varphi}} \simeq \frac{\M^2}{2}\Bigl(\frac{V'}{V}\Bigr)^2 + \frac{\varphi}{4}\frac{V'}{V} + \frac{1}{4} - \M^2 \frac{V''}{V},
  \label{sp10}
\end{equation}
where in the formal expansion higher order terms have been neglected.

To simplify comparison with scalar field inflation, it is most natural use the following set of scalar field slow roll parameters
\begin{equation}
  \epsilon = \frac{\M^2}{2}\Bigl(\frac{V'}{V}\Bigr)^2, \qquad
  \delta = \epsilon - \M^2 \frac{V''}{V}.
  \label{sp11}
\end{equation}
Inserting these parameter into Eqs.~(\ref{sp9}) and~(\ref{sp10}) leads to
\begin{eqnarray}
  \mathcal{E} = -\frac{\dot{H}}{H^2} \simeq \epsilon - \frac{\varphi}{4}\frac{V'}{V},
  \label{sp12}\\
  \Delta = \frac{1}{H}\frac{\ddot{\varphi}}{\dot{\varphi}} \simeq \delta + \frac{1}{4} + \frac{\varphi}{4}\frac{V'}{V},
  \label{sp13}
\end{eqnarray}
where the additional term $1/4$ should be noted, see also~\cite{Boehmer:2007ut}. The left-hand side of Eq.~(\ref{sp12}) is motivated by the definition of inflation $\ddot{a}/a = \dot{H} + H^2 > 0$, which the yields $-\dot{H}/H^2<1$. In scalar field inflation one can therefore naturally define the end of inflation by the condition $\epsilon = 1$. In spinor inflation the additional term $\varphi/4\, V'/V$, depending on its sign, may stop inflation even if $\epsilon \ll 1$ or spinflation may continue if $\epsilon = 1$. 

\section{Exponential inflation}
\label{sec:exact}

In this section the non-approximated and the slow-roll approximated equations of motion are analyzed. In scalar field equation the slow-roll approximation suffices for the analysis of most models.

Consider the potential
\begin{equation}
  \frac{V(\varphi)}{\M^2} = 3 q^2 + \frac{q^2}{4}\frac{\varphi^2}{\M^2},
  \label{ex1}
\end{equation}
where $q$ is a constant of mass dimension one. The two Einstein field equations~(\ref{eq10}) and~(\ref{eq11}) are solved by
\begin{equation}
  a = a_0 \exp(q t),\qquad \varphi = \varphi_0 \exp(\pm q t /2).
  \label{ex2}
\end{equation}
This spinflation model is characterized by a constant Hubble parameter $H=q$. Moreover, the term $\varphi/\dot{\varphi}$ is constant and therefore all terms of interest in~(\ref{sp4}) and~(\ref{sp5}) are constants. If this potential is used to solve the slow-roll approximated equations of motion~(\ref{sp4})--(\ref{sp5}), one finds the same solution.

The number of $e$-foldings for this exponential inflation model is $N = q (t_{\rm end} - t)$ and therefore we find $q\, t_{\rm end} = 60$ which relates the time when inflation ends to the parameter $q$ that parameterizes the potential. Note that the phase of accelerated expansion never ends since $\ddot{a}/a = q^2 >0$. Moreover, the inflation condition $V(\varphi) > \dot{\varphi}^2$ yields $q \M > 0$ which is also always satisfied, this is true for both signs in Eq.~(\ref{ex2}). The parameters $\mathcal{E}$ and $\Delta$ are given by
\begin{equation}
  \mathcal{E} = 0, \qquad \Delta = \pm \frac{1}{2}.
  \label{ex2a}
\end{equation}
Having obtained these values, we can now consider Eqs.~(\ref{sp12}) and~(\ref{sp13}) for consistency. The value $\mathcal{E}=0$ implies that $\epsilon \simeq \varphi/4\, V'/V$, which leads to a very small value of the field. Indeed, for $\varphi = 0$ this condition holds exactly. On the other hand, $\Delta=1/2$ holds exactly if $\varphi = \pm 2 \sqrt[4]{15}\M$.

Next, let us analyze the additional terms present in Eq.~(\ref{sp5}) which are given by
\begin{eqnarray}
  \frac{\varphi}{\dot{\varphi}}\,\frac{\dot{a}}{a} = \pm 2\qquad
  \frac{\varphi^2}{\dot{\varphi}^2}\,\frac{\ddot{a}}{a} = 4.
  \label{ex2b}
\end{eqnarray}
Taking into account the factor of $1/4$ we conclude that one cannot neglect the additional terms in the square bracket of Eq.~(\ref{sp5}) as these are of the order unity and hence spinflation shows genuine differences from scalar field inflation.

\section{Cosmological perturbations and power spectrum} 
\label{sec:pert}

In cosmological perturbation theory it is most convenient to work in conformal time defined by $\eta = \int a^{-1}(t')dt'$. The derivative with respect to $\eta$ will be denoted by a prime. The metric now takes the form
\begin{equation}
  ds^2 = a(\eta)^2 (d\eta^2 - dx^2 - dy^2 - dz^2).
  \label{flrwconf}
\end{equation}
Since we wish to solve the perturbation equation explicitely for the exponential inflation model, we mention its form in conformal time
\begin{eqnarray}
  a(\eta) = -\frac{1}{q} \frac{1}{\eta},\\
  \varphi(\eta) = \sqrt{\frac{\bar{\varphi}}{\eta}}\quad\mbox{case\ I},\qquad
  \varphi(\eta) = \sqrt{\bar{\varphi}^3 \eta},\quad\mbox{case\ II},
  \label{cos2}
\end{eqnarray}
depending on the sign in Eq.~(\ref{ex2}), $+$ and $-$ refer to cases I and II respectively. The constant $\bar{\varphi}$ has mass dimension one in both cases.

It is well known in scalar field inflation that one can neglect the metric perturbation when deriving the wave equation of the field perturbations~\cite{Dodelson:2003}. This is justified for two reasons. Firstly, for modes well inside the horizon the metric perturbation $\mathbf{\Psi}$ is negligibly small. Secondly, when working in spatially flat slicing gauge, the spatial part of the metric is indeed unperturbed. Since we are primarily interested in qualitative statements about spinflation, we will neglect the metric perturbations for the time being.

Let us consider small perturbation about the homogeneous and isotropic spinflaton field 
\begin{equation}
  \varphi = \varphi_0(\eta) + \delta\varphi(\eta,\vec{x}),
  \label{cos3}
\end{equation}
we will keep first order terms in $\delta\varphi$ only, $\vec{x}$ denotes the three spatial coordinates. It should be noted that we neglect perturbations of the spinorial parts $\xi$ and $\zeta$ of the spinflaton field as a first approximation. At the moment it is unclear whether this is fully justified. Due to the more involved equation of motion of the spinor field~(\ref{eq11}), one expects a rather complicated perturbation equation. Indeed, this wave equation is given by
\begin{eqnarray}
  \delta\varphi'' + 2 a H \delta\varphi' + k^2\delta\varphi + 
  V_{\varphi\varphi}(\varphi_0)\, a^2\, \delta\varphi
  \nonumber \\[1ex]
  + \frac{3}{4}\frac{\varphi_0^2}{\varphi_0'^2} (a H^2 + H')\, a^2 H\, \delta\varphi'
  \nonumber \\[1ex]
  - \frac{3}{2}\frac{\varphi_0}{\varphi_0'} (a H^2 + H')\, a^2 H\, \delta\varphi 
  -\frac{3}{4} H^2 a^2\, \delta\varphi = 0.
  \label{cos4}
\end{eqnarray} 
where the first line contains the terms of standard scalar field inflation and $k$ is the comoving wavenumber. Evidently, this wave equation has an involved coupling to the background dynamics.

Assuming the background evolution to be described by the exponential inflation model of Section~\ref{sec:exact}, the wave equations~(\ref{cos4}) simplify considerably and one arrives at
\begin{eqnarray}
  \delta\varphi'' - \frac{5}{\eta}\delta\varphi' + 
  \left(k^2 -\frac{13}{4}\frac{1}{\eta^2}\right)\delta\varphi = 0,
  \quad\mbox{case\ I},
  \label{cos5}\\[1ex]
  \delta\varphi'' - \frac{5}{\eta}\delta\varphi' + 
  \left(k^2 +\frac{11}{4}\frac{1}{\eta^2}\right)\delta\varphi = 0,
  \quad\mbox{case\ II},
  \label{cos5a}
\end{eqnarray}
These equations can be brought into the standard form of a harmonic oscillator by eliminating the term with $\delta\varphi'$. Introducing $v=(q\eta)^{-5/2}\delta\varphi\,[=i a^{5/2}\delta\varphi]$ we obtain
\begin{eqnarray}
  v'' + \left(k^2-\frac{12}{\eta^2} \right)v = 0,\quad\mbox{case\ I},\\[1ex]
  v'' + \left(k^2-\frac{6}{\eta^2} \right)v = 0,\quad\mbox{case\ II},
\end{eqnarray}
which can be solved analytically in terms of trigonometric functions. We impose the usual Minkowski vacuum ground state
\begin{equation}
  v \rightarrow \frac{e^{-ik\eta}}{\sqrt{2k}},
  \label{cos7}
\end{equation}
in the limit $k\eta\rightarrow -\infty$, the asymptotic past. This fixes the two constants of integration. Therefore, the properly normalized solutions are given by
\begin{eqnarray}
  v = \frac{e^{-ik\eta}}{\sqrt{2k}}
  \left(1-\frac{i\,6}{k\eta}-\frac{15}{k^2\eta^2}+\frac{i\,15}{k^3\eta^3}\right),
  \quad\mbox{case\ I},
  \label{cos8}\\[1ex]
  v = \frac{e^{-ik\eta}}{\sqrt{2k}}
  \left(1-\frac{i\,3}{k\eta}-\frac{3}{k^2\eta^2}\right),
  \quad\mbox{case\ II},
  \label{cos8a}
\end{eqnarray}
where we immediately find that the large scale power spectrum will be different for the two cases. 

Since we wish to directly compare scalar field inflation with spinor inflation, we define the power spectrum of the spinor field perturbations according to
\begin{equation}
  \mathcal{P}_{\delta\varphi} = \frac{4\pi k^3}{(2\pi)^3}\frac{|v|^2}{a^{5/2}}.
  \label{cos9}
\end{equation}
Now we can find the power spectrum on small and large scales. On small scales the comoving wave number $k$ satisfies $k \gg aH$, therefore the two resulting power spectra are in fact equal and become
\begin{equation}
  \mathcal{P}_{\delta\varphi} = \left(\frac{k}{2\pi\, a^{5/4}}\right)^2.
  \label{cos10}
\end{equation}
Besides the slightly different scaling in time the small scale power spectrum coincides with the prediction of scalar field inflation, where 
\begin{equation}
  \mathcal{P}_{\delta\phi} = \left(\frac{k}{2\pi\, a}\right)^2,
  \label{cos10a}
\end{equation}
and on large scales where $k \ll aH$ we find
\begin{eqnarray}
  \mathcal{P}_{\delta\varphi} = \left(\frac{H}{2\pi}\right)^2
  \left(\frac{15\, H^2}{k^2 a^{7/4}}\right)^2 =
  \left(\frac{H}{2\pi}\right)^2 15^2 \left(\frac{H}{k}\right)^{15/2} |k\eta|^{7/2},
  \quad\mbox{case\ I},
  \label{cos11}\\[1ex]
  \mathcal{P}_{\delta\varphi} = \left(\frac{H}{2\pi}\right)^2 
  \left(\frac{3\, H}{k a^{3/4}}\right)^2 =
  \left(\frac{H}{2\pi}\right)^2 3^2 \left(\frac{H}{k}\right)^{7/2} |k\eta|^{3/2},
  \quad\mbox{case\ II}.
  \label{cos11a}
\end{eqnarray}
The first factors on the left are those that resemble the massless field case. Therefore, we are lead to conclude that on large scales the spinflaton field perturbations have more power than the scalar field perturbations.

\section{Conclusions}
\label{sec:concl}

The main difference between spinflation and scalar field inflation lies in the additional Hubble parameter terms in the energy density and pressure of the dark spinor field. These terms seem to change the dynamical behavior of the theory significantly but also offer a new and well motivated inflation model. We formulated the theory of spinflation in such a manner that the theory looks closely analogous to conventional inflation. We found that dark spinor driven inflation shares many of the desired properties of the successful scalar field inflation model but also shows genuine differences which make the spinflation theory very interesting. In principle it should now be possible to thoroughly investigate spinflation models for various potentials and compare with observations. In view of the apparent alignment of the cosmic microwave background multipoles on large scales~\cite{Eriksen:2003db,Copi:2003kt,deOliveiraCosta:2003pu,Land:2005ad}, alternative inflation models are indeed required assuming that this apparent alignment is not a data analysis effect.

As a possible next step one should take into account the metric perturbations~\cite{Bardeen:1980kt,Kodama:1985bj,Mukhanov:1990me}. In spatially flat gauge, the comoving curvature perturbation is defined by
\begin{equation}
  \mathcal{R} = \frac{2}{3}\rho_\varphi\frac{H^{-1}\dot{\Phi}+\Phi}{\rho_\varphi+P_\varphi}+\Phi.
  \label{cos12}
\end{equation}
For the exponential spinflation model discussed so far, one immediately encounters the problem that the term $\rho_\varphi+P_\varphi$ for the exact solution~(\ref{ex2}) vanishes. Therefore, dividing by the term $\rho_\varphi+P_\varphi$ for that particular model is not allowed and one has to investigate the perturbations starting from the perturbed Einstein field equations directly. Therefore, in order to find the spectral index from the curvature perturbation power spectrum, additional investigations are required. Since the spinflaton field has a more complicated coupling to the geometry, it is expected that also the complete perturbation equations will have a more involved structure. Moreover, in principle one could consider perturbations of the spinorial part of the field that are compatible with homogeneity and isotropy. This leads to additional structure in the resulting equations and may also have an effect on the large scale power spectrum.

Within the exponential inflation model there is no natural end for spinflation, see Eq.~(\ref{sp12}) and the subsequent discussion. Therefore, it is of interest to find further exact solutions of the equations of motion based on other potentials and to analyze under which conditions spinflation ends. Another important feature of Eq.~(\ref{eq11}) should be remarked. The additional terms containing the Hubble parameter effectively contribute to the potential energy of the field. This allows for situations where the field in not oscillating around its minimum and therefore is not decaying. One can speculate that this could yield interesting scenarios where the spinflaton field firstly drives an epoch of accelerated expansion and later constitutes to the cosmological dark matter. To conclude, dark spinor inflation is an intriguing new inflation model with many features yet to be discovered. 

\acknowledgments
I thank Roy Maartens and Jochen Weller for valuable comments on the manuscript.


\end{document}